\documentclass[twocolumn,showpacs,preprintnumbers]{revtex4}
\usepackage{graphicx}
\usepackage{dcolumn}
\usepackage{bm}
\begin{document}
\newcommand{\ds}{\displaystyle}
\newcommand{\be}{\begin{equation}}
\newcommand{\en}{\end{equation}}
\newcommand{\bea}{\begin{eqnarray}}
\newcommand{\ena}{\end{eqnarray}}
\topmargin 0cm
\title{Duality mappings within three-dimensional nonlinear electrodynamics}
\author{Mauricio Cataldo}
\altaffiliation{mcataldo@ubiobio.cl}
\affiliation{Departamento de F\'\i sica, Facultad de Ciencias,
Universidad del B\'\i o-B\'\i o, Avenida Collao 1202, Casilla 5-C,
Concepci\'on, Chile.\\}
\date{\today}
\begin{abstract}
{\bf {Abstract:}} In three-dimensional Einstein-Maxwell gravity
the electrostatic Ba\~nados-Teitelboim-Zanelli solution and the
magnetostatic Hirschmann-Welch solution are connected by a duality
mapping. Here we point out that a similar duality mapping exists
among circularly symmetric electrostatic and magnetostatic
spacetimes, and electric and magnetic stationary solutions, for a
nonlinear electrodynamics coupled to three-dimensional Einstein
gravity.

\vspace{0.5cm} 
\end{abstract}
\smallskip\
\maketitle \preprint{APS/123-QED}

\section{Introduction}
In three-dimensional Einstein-Maxwell gravity exist the electric
and magnetic analogs to the Reissner-Norsdtr\"{o}m-Kottler
solution. They are the Ba\~{n}ados-Teitelboim-Zanelli (BTZ) black
hole~\cite{Teitelboim}
\begin{eqnarray}
\label{BTZ} ds^{2} &=&-(r^{2}/l^{2}-M-q_{_{e}}^{2}\ln
r)d\tilde{t}^{2}
\\
&&+\frac{dr^{2}}{(r^{2}/l^{2}-M-q_{_{e}}^{2} \ln
r)}+r^{2}d\tilde{\theta}^{2}, \nonumber
\end{eqnarray}
where the radial electric field is given by $E=q_{e}/r$, and the
Hirschmann-Welch (HW) magnetic solution~\cite{HW}
\begin{eqnarray}
\label{HW} ds^{2} =-(r^{2}/l^{2}-M)d\tilde{t}^{2} \nonumber \\
+\frac{r^{2}dr^{2}}{(r^{2}/l^{2}-M)(r^{2}+q_{_{m}}^{2}\ln (r^{2}/l^{2}-M))}%
\\ \nonumber +(r^{2}+q_{_{m}}^{2} \ln (r^{2}/l^{2}-M))d\tilde{\theta}^{2},
\nonumber
\end{eqnarray}
where the pseudoscalar magnetic field is given by $B=q_{m}/r$.
Here $\Lambda =-1/l^{2}<0$, $q_{_{e}}$ is the electric charge,
$q_{_{m}}$ magnetic charge and $M$ is the mass of the BTZ black
hole. Both solutions are asymptotically anti-de Sitter like the
Reissner-Norsdtr\"{o}m-Kottler solution for $\Lambda <0.$

In~\cite{Cataldo1} it is shown that there exist a duality mapping
which connects the BTZ black hole and the HW solution. This
mapping is given by
\[
q_{_{e}}\longrightarrow i\,q_{_{m}},\,\,\,\tilde{t}\longrightarrow
i\,\theta ,\,\,\,\tilde{\theta}\longrightarrow i\,t.
\]
In this case the BTZ black hole takes the form
\begin{eqnarray}
\label{HWmio} ds^{2} &=&-r^2 d\tilde{t}^{2}
+\frac{dr^{2}}{(r^{2}/l^{2}-M+q_{_{m}}^{2}\ln r)}  \nonumber \\
&&+(r^{2}/l^{2}-M+q_{_{m}}^{2} \ln r) d\tilde{\theta}^{2}.
\end{eqnarray}
It could be seen, by introducing a new coordinate $r$, that the
metric~(\ref{HWmio}) becomes~(\ref{HW}).

This property is related to the fact that in (2+1)-dimensions the
Maxwell tensor has only three independent components, two for the
vector electric field and one for the scalar field. This means
that the duality transformation which takes place for the
Reissner-Norsdtr\"{o}m-Kottler black hole do not take place now.
The reason is that in (3+1)-dimensions the Maxwell tensor and its
dual are two-forms and in (2+1)-dimensions the Maxwell tensor is a
two-form and its dual is a one form *$F_{a}=1/2\epsilon
_{abc}F^{bc}$. This implies that the electric and magnetic
counterparts are quite different.

In this letter we point out that a similar duality mapping exist
among circularly symmetric electrostatic and magnetostatic
spacetimes for a nonlinear electrodynamics coupled to
three-dimensional Einstein gravity.

The letter is organized as follows. The next section deals with
duality mappings among electrostatic and magnetostatic
Einstein-nonlinear Maxwell spacetimes and we discuss some specific
examples of nonlinear magnetic solutions. In Sec. III we discuss
stationary Einstein-nonlinear Maxwell three-dimensional
spacetimes. The electric and magnetic spinning cases are discussed
separately. Finally, in Sec. IV some properties of these spinning
nonlinear three-dimensional spacetimes are discussed.
\section{Duality mappings within static Einstein-nonlinear Electrodynamics}
In (2+1)-dimensions, electromagnetic theories can be
constructed from Lagrangians (without higher derivatives)
depending on a single (non-vanishing) invariant
$F=\frac{1}{4}F_{ab}F^{ab}$, which can be expressed in terms of
the electric (vector) and magnetic (pseudoscalar) fields: in a
lorentzian frame, for an observer moving with the 3-velocity
$v^{a}$, the electric and the magnetic fields are correspondingly
defined as
\[
E_{a}=F_{ab}v^{b},\,\,\,B=\frac{1}{2}\epsilon _{abc}F_{bc}v^{a},
\]
where latin indices run the values $0,1,2$ and $\epsilon _{abc}$
is the totally anti-symmetric Levi-Civita symbol with $\epsilon
_{012}=1$, usually the $v^{a}$ is oriented along the time
coordinate, i.e, $v^{a}=\delta _{t}^{a}$, with such a choice
\[
E_{a}=F_{a0},\,\,\,B=F_{12}.
\]
Thus the invariant can be expressed by
\[
F\equiv \frac{1}{4}F^{ab}F_{ab}=%
\frac{1}{2}(B^{2}-E^{2}).
\]

In general one can construct a (2+1)-Einstein theory coupled to a
nonlinear electrodynamics starting from the action
\[
S=\int \sqrt{-g}\left( \frac{1}{16\pi }(R-2\Lambda )+L(F)\right)
\,d^{3}x,
\]
with the electromagnetic Lagrangian $L(F)$ unspecified explicitly
at this stage; physically one requires the Lagrangian to coincide
with the Maxwell one at small values of the electromagnetic
fields, $L(F)_{Maxwell}=-F/{4\pi } $. We are using units in which
$c=G=1$, because of the ambiguity in the definition of the
gravitational constant (there is not newtonian
gravitational limit in (2+1)-dimensions) we prefer to maintain the factor $%
1/16\pi $ in the action to keep the parallelism with
(3+1)-gravity. Varying this action with respect to gravitational
field gives the Einstein equations
\begin{eqnarray*}
G_{ab}+\Lambda g_{ab}=8\pi T_{ab},
\end{eqnarray*}
where
\begin{eqnarray}
\label{tensor E-M} T_{ab}=g_{ab}L(F)-F_{ac}F_{b}^{\,\,c}L_{_{,F}},
\end{eqnarray}
while the variation with respect to the electromagnetic potential
$A_{a}$ entering in $F_{ab}=A_{b,a}-A_{a,b}$, yields the
electromagnetic field equations
\begin{eqnarray} \label{ecdecampoelectricas}
\nabla _{a}\left( F^{ab}L_{_{,F}}\right) =0,
\end{eqnarray}
where $L_{_{,F}}$ denotes the derivative of $L(F)$ with respect to
$F$.

Concrete solutions to the dynamical equations we shall seek in the
form
\begin{eqnarray} \label{metrica}
ds^2=-e^{2 \alpha(r)}dt^2+e^{2 \beta(r)}dr^2+e^{2
\gamma(r)}d\phi^2,
\end{eqnarray}
where $\alpha$, $\beta$ and $\gamma$ are unknown functions of the
variable $r$. We have chosen this form because if the structural
functions are given by $e^{2 \alpha(r)} = e^{-2\beta(r)}$ and
$e^{2 \gamma(r)}=r^2$ a magnetic field can not be
obtained~\cite{Cataldo2}.

The general form of the electromagnetic field tensor which shares
the static circularly symmetric spacetime is given by
\begin{eqnarray} \label{tensordeMaxwell}
F_{_{M}}= E_{r} \, dr \, \wedge \,dt + B \, dr \, \wedge
\,d\theta.
\end{eqnarray}
This gives us the following invariant $F$ (from now on $E_r\equiv
E$):
\begin{eqnarray} \label{invariante}
F=\frac{1}{4} \, F_{ab}F^{ab}= \frac{1}{2} \, \left(-E^2e^{-2
\alpha-2\beta}+B^2e^{-2\beta-2\gamma}\right).
\end{eqnarray}

Thus from~(\ref{tensor E-M}) we have for nonvanishing components
\begin{eqnarray}
T_{00}&=&-e^{2 \alpha} L(F)-E^2 \, e^{-2\beta} L_{,_{F}} \nonumber
\\ T_{11}&=&e^{2 \beta} L(F)+(E^2 \, e^{-2\alpha}-B^2
e^{-2\gamma}) L_{,_{F}} \nonumber \\ T_{22}&=&e^{2 \gamma}
L(F)-B^2 \, e^{-2\beta} L_{,_{F}} \nonumber
\\ T_{02}&=&-E B e^{-2\beta} L_{,_{F}}.
\end{eqnarray}
Then the Einstein equations are
\begin{eqnarray}  \label{cero-cero}
e^{- 2 \beta} \left( \gamma {}^{\prime} \beta {}^{\prime} - \gamma
{}^{\prime \prime}- \gamma {}^{\prime}{}^{2}\right) = \nonumber \\
\Lambda + \kappa \left[-L(F)-E^2 \, e^{-2\alpha-2\beta} L_{,_{F}}
\right],
\end{eqnarray}

\begin{eqnarray}  \label{uno-uno}
\alpha\, ^{\prime}\, \gamma\, ^{\prime}\,e^{- 2\, \beta}= -\Lambda
\nonumber \\  + \kappa \left[L(F)+(E^2 \, e^{-2\alpha-2\beta}- B^2
e^{-2\beta-2\gamma}) L_{,_{F}} \right],
\end{eqnarray}

\begin{eqnarray}  \label{dos-dos}
e^{- 2 \beta} \left( \alpha {}^{\prime} \beta {}^{\prime}- \alpha
{}^{\prime\prime}- \alpha {}^{\prime}{}^{2} \right) = \nonumber
\\ \Lambda + \kappa \left[-L(F)+B^2 \, e^{-2\beta-2\gamma}
L_{,_{F}} \right]
\end{eqnarray}
\begin{eqnarray}  \label{cero-dos}
E B  L_{,_{F}}= 0,
\end{eqnarray}
where $\kappa=8 \pi$ and the prime denotes the differentiation
$d/dr$. From the electromagnetic fields
equations~(\ref{ecdecampoelectricas}) we arrive at
\begin{eqnarray} \label{condicionelectrica}
L_{,_{F}} E=C_1 e^{\alpha+\beta-\gamma}
\end{eqnarray}
and
\begin{eqnarray} \label{condicionmagnetica}
L_{,_{F}} B=C_2 e^{-\alpha+\beta+\gamma},
\end{eqnarray}
where $C_1$ and $C_2$ are constants of integration.

Notice that from the Einstein equations we conclude that $E$ or
$B$ must be zero, i.e. we cannot have any superposition of these
nonlinear electric and magnetic fields in any three-dimensional
static spacetime, given by diagonal metric~(\ref{metrica}), just
like in the linear case.

On the other hand, from the invariant~(\ref{invariante}) and the
equations~(\ref{cero-cero})--(\ref{condicionmagnetica}) we see
that the self-consistent equations are invariant under the duality
mapping
\begin{eqnarray} \label{mapeodual}
\alpha \longrightarrow \gamma, \,\,\,\,\, \gamma \longrightarrow
\alpha \nonumber \\ E \longrightarrow i \, B, \,\,\,\,\, B
\longrightarrow i \, E.
\end{eqnarray}
Thus we conclude that if we have a nonlinear electric static
solution, then one can obtain the nonlinear magnetic one by making
the dual mapping~(\ref{mapeodual}).

For solving the field equations we must set $E=0$ (for magnetic
solutions) or $B=0$ (for electric solutions). In this case the
magnetic solutions can be written using the result of
Ref.~\cite{Cataldo2}, where it was obtained the general solution
for the static (2+1)-spacetime for any nonlinear Lagrangian
depending on the electric field (i.e. $B=0$). These electric
solutions have the form~(\ref{metrica}) with structural functions
given by
\begin{eqnarray}\label{metrica electrostatica}
e^{2 \alpha}=e^{-2 \beta}&=&- M -\Lambda r^{2} + 16 \pi \int
r[L+E^2 L_{,_{F}}]dr, \nonumber \\ e^{2 \gamma}&=&r^2
\end{eqnarray}
and, where $M$ is a constant of integration. Then using the dual
mapping~(\ref{mapeodual}) we obtain the nonlinear magnetic
spacetime given by~(\ref{metrica}), where
\begin{eqnarray} \label{solucionprefinal}
e^{2 \alpha}&=& r^2, \nonumber \\ e^{-2 \beta}=e^{2 \gamma}&=& - M
-\Lambda r^{2} + 16 \pi \int r[L-B^2 L_{,_{F}}]dr. \nonumber \\
\end{eqnarray}
In this case the invariant is $F=B^2/2$. Then
from~(\ref{condicionmagnetica}) we have that
\begin{eqnarray}
B(r) L_{,_{F}}=\frac{Const}{r}
\end{eqnarray}
and we can write
\begin{eqnarray}
L_{,r}=\frac{Const}{r} \, B_{,r}.
\end{eqnarray}
In order to obtain the Maxwell limit we must set $Const=-q/4\pi$
and from~(\ref{solucionprefinal}) we obtain the solution in terms
of the magnetic field $B(r)$:
\begin{eqnarray} \label{solucionfinal}
e^{2 \alpha}&=& r^2, \nonumber \\ e^{-2 \beta}=e^{2 \gamma}&=& - M
-\Lambda r^{2} - 4q\int \left[r\int \frac{B_{,r}}{r}dr-B
\right]dr. \nonumber
\\
\end{eqnarray}
It is easy to check that for Maxwell case, i.e. when $L=-F/4\pi$,
we have $B=q/r$ and then from~(\ref{solucionfinal}) we obtain the
HW magnetic solution~(\ref{HWmio}).

\subsection{Some applications}
We present now various particular examples of nonlinear
magnetic solutions.

One interesting application of our results arises in the
Born-Infeld electrodynamics~\cite{Cataldo3}. In this case the
Born-Infeld three-dimensional electric spacetime is given by
\begin{eqnarray}\label{Born-Infeld B-H}
e^{2\alpha}&=&e^{-2\beta} = - M - (\Lambda - 2 b^{2}) r^{2} -2
b^{2}r \sqrt{r^{2}+ q_e^{2}/b^{2}}  \nonumber \\ &-& 2 q_e^{2} \ln
(r +\sqrt{r^{2}+q_e^{2}/b^{2}} ), \,\,\,\,\, e^{2 \gamma}=r^2,
\end{eqnarray}
and the Lagrangian and the electric field are given by
\begin{eqnarray}
L(F)=-\frac{b^{2}}{4 \pi} \left (\sqrt{1+2 \frac{F}{b^{2}}}-1
\right)= \nonumber \\ -\frac{b^2}{4 \pi} \left( \frac{r}{\sqrt{r^2
+ q_e^2/b^2}} -1 \right), \nonumber \\ E(r)= \frac{q_e}{\sqrt{r^2
+ q_e^2/b^2}}, \hspace{2 cm}
\end{eqnarray}
where $b$ is the Born-Infeld parameter.

Applying the dual mapping in the form $\alpha \rightleftharpoons
\gamma$ and $q_e=i \, q_m$ we obtain for the Born-Infeld magnetic
version
\begin{eqnarray}\label{magnetic B-I} e^{2 \alpha}=r^2, \,\,\,\,\,
e^{-2\beta}=e^{2\gamma} = - M - (\Lambda - 2 b^{2}) r^{2}
\nonumber \\-2 b^{2}r \sqrt{r^{2}- q_m^{2}/b^{2}} + 2 q_m^{2} \ln
(r +\sqrt{r^{2}-q_m^{2}/b^{2}} ),
\end{eqnarray}
and the Lagrangian and the magnetic field are given by
\begin{eqnarray}
L(F)=-\frac{b^{2}}{4 \pi} \left (\sqrt{1+2 \frac{F}{b^{2}}}-1
\right)= \nonumber \\ -\frac{b^2}{4 \pi} \left( \frac{r}{\sqrt{r^2
- q_m^2/b^2}} -1 \right), \nonumber \\ B(r)= \frac{q_m}{\sqrt{r^2
- q_m^2/b^2}}. \hspace{2 cm}
\end{eqnarray}
Other interesting example is the magnetic solution obtained from
the electric (2+1)-regular black hole solution~\cite{Cataldo2}
\begin{eqnarray}\label{regular B-H}
e^{2 \alpha}=e^{-2 \beta}&=&- M - \Lambda r^2 - q_e^2 \ln
(r^2+a^2), \nonumber \\ e^{2 \gamma}&=&r^2
\end{eqnarray}
where $M$, $a$, $q_e$ and $\Lambda$ are free parameters. The
Lagrangian and the electric field are given by
\begin{eqnarray}
L(r)= \frac{q_e^2}{8 \pi} \, \frac{(r^2 - a^2)}{(r^2 + a^2)^2},
\nonumber \\ E(r)= \frac{q_e r^3}{(r^2+a^2)^2}.
\end{eqnarray}
In this case the metric, the electric field and all invariants
behave regularly for all values of $r$ and then this solution is
curvature regular everywhere. Applying the dual mapping in the
form $\alpha \rightleftharpoons \gamma$ and $q_e=i \, q_m$ we
obtain for the magnetic version
\begin{eqnarray} \label{regular magnetic}
e^{2 \alpha}&=&r^2, \nonumber \\ e^{-2 \beta}=e^{2 \gamma}&=&- M -
\Lambda r^2 + q_m^2 \ln (r^2+a^2)
\end{eqnarray}
where $M$, $a$, $q_m$ and $\Lambda$ are free parameters. The
Lagrangian and the electric field are given by
\begin{eqnarray}
L(r)= -\frac{q_m^2}{8 \pi} \, \frac{(r^2 - a^2)}{(r^2 + a^2)^2},
\nonumber \\ B(r)= \frac{q_m r^3}{(r^2+a^2)^2}.
\end{eqnarray}
The last example is the (2+1)-dimensional magnetic solution with a
Coulomb-like field. It is obtained from the (2+1)-dimensional
black hole with Coulomb-like electric field. In this case the
energy momentum tensor~(\ref{tensor E-M}) is traceless. The
electrostatic spacetime is given by
\begin{eqnarray}\label{Coulomb-like elec}
e^{2 \alpha}=e^{-2\beta}&=&- M - \Lambda r^{2} + \frac{4 q_e^2}{3
r}, \nonumber \\ e^{2 \gamma}&=&r^2,
\end{eqnarray}
where $M$ is a constant of integration related to the mass at
infinity ~\cite{Cataldo4}. The Lagrangian and the electric field
are given by
\begin{eqnarray}
L = \frac{\sqrt{q_e}}{6 \pi} \, |F|^{3/4}&=&\frac{q_e^2}{6 \pi
r^3} , \nonumber \\ E(r)&=&\frac{q_e}{r^2},
\end{eqnarray}
where the bars denote absolute value. Applying the dual mapping we
obtain
\begin{eqnarray}\label{magnetic coulomb}
e^{2 \alpha}&=&r^2, \nonumber \\ e^{-2\beta}=e^{2 \gamma}&=&- M -
\Lambda r^{2} - \frac{4 q_m^2}{3 r}.
\end{eqnarray}
The Lagrangian and the magnetic field are given by
\begin{eqnarray}
L &=& -\frac{q_m^2}{6 \pi r^3} , \nonumber
\\ B(r)&=&\frac{q_m}{r^2}.
\end{eqnarray}

Thus, for any three-dimensional static nonlinear electrodynamics
whose Lagrangian $L$ depends on a single invariant
$F=F_{ab}F^{ab}/4$, and $F_{ab}$ is given
by~(\ref{tensordeMaxwell}), exist a dual mapping~(\ref{mapeodual})
between electrostatic and magnetostatic counterparts.

\section{Stationary Einstein-nonlinear Maxwell three-dimensional spacetimes}
Now we shall consider the generalization of these
Einstein-nonlinear Maxwell three-dimensional static fields to
include the angular momentum $J$. In this case one requires
invariance under spatial rotations and time translations, then the
metric can be written in the following general form
\begin{eqnarray} \label{metrica rotante}
ds^2=-(N^0(r))^2 \, dt^2+\frac{dR^2}{G^2(r)}
+R^2(r)(N^{\theta}(r)\, dt+d\theta)^2.\nonumber \\
\end{eqnarray}
One could find any rotating solution by solving the
Einstein-nonlinear Maxwell equations using the electromagnetic
field tensor~(\ref{tensordeMaxwell}) and the metric~(\ref{metrica
rotante}). However, in order to do this we shall find the
solutions by means of a rotational Lorentz boost. This kind of
transformations can generate from seed static spacetimes new
metrics because they are not permitted global coordinate
transformations. This procedure of generating new solutions from
seed ones has been quite useful in general relativity. A
gravitational field may be locally the same but globally distinct
due to differences in the topology of their underlying manifolds.
Globally stationary but locally static gravitational fields
provide a good example of this fact ~\cite{Stachel}.

In view that for the considered Einstein-nonlinear Maxwell problem
there are two branches of solutions, electrostatic and
magnetostatic one, we must apply the Lorentz boost separately.

\subsection{Stationary electric nonlinear spacetimes}
We first shall consider the electrostatic nonlinear spacetimes. In
the linear case, i.e. Maxwell electrodynamics, the extension to
consider the rotating electrically charged BTZ black hole has been
derived in references~\cite{Clement1,Clement2,Teitelboim1}. For
obtaining the stationary electric spacetime we shall follow
essentially the procedure of the mentioned above
Ref.~\cite{Teitelboim1}. The ``rotation boost" may be written as
the following transformation~\cite{Teitelboim1}
\begin{eqnarray} \label{boostt}
\tilde{t}&=&\frac{t-\omega \theta }{\sqrt{1-\omega ^{2}}},
\nonumber \\ \widetilde{\theta }&=&\frac{\theta -\omega
t}{\sqrt{1-\omega ^{2}}}. \label{boost}
\end{eqnarray}
Here the $\tilde{t}$ and $\widetilde{\varphi}$ are the old time
and angular coordinates, respectively, used to write the static
form of the metric.

In this case the new metric may be written in the
form~(\ref{metrica rotante}), where $N^0(r)=N(r)G(r)$ and
\begin{eqnarray}
R^{2} &=&\frac{r^{2}-\omega ^{2}f^{2}}{1-\omega ^{2}},
\label{R2(f)} \\ G^{2} &=&\left( \frac{dR}{dr}\right) ^{2}f^{2},
\label{F2} \\ N &=&\frac{r}{R}\left( \frac{dr}{dR}\right) =
\frac{d(r^2)}{d(R^2)},  \label{N} \\ N^{\theta } &=&\frac{\omega
(f^{2}-r^{2})}{(1-\omega ^{2})R^{2}}. \label{Nphi}
\end{eqnarray}
If one starts with the charged nonrotating electrostatic black
hole~(\ref{metrica electrostatica}), then $f(r)=e^{\alpha(r)}$.

We shall consider asymptotically anti de Sitter solution, i.e. the
cosmological constant is negative and is related to the
cosmological length $l$ by $\Lambda=-1/l^2$. Now on this length
will be set equal to unity. Thus we have
\begin{eqnarray}\label{felectrica}
f^2(r)=- M + r^{2} + 16 \pi \int r[L+E^2 L_{,_{F}}]dr.
\end{eqnarray}
Applying the boost~(\ref{boost}) to the solution~(\ref{metrica
electrostatica}) we obtain
\begin{eqnarray}
R^2=r^2+\frac{\omega^2}{1-\omega^2} \left(M - 16 \pi \int r[L+E^2
L_{,_{F}}]dr \right), \label{R2Q}
\end{eqnarray}
\begin{eqnarray}
G^2=\frac{\left(r-\frac{8 \omega^2\pi r (L+E^2
L_{,_{F}})}{1-\omega^2} \right)^2}{R^2} \nonumber \\ \times
\left(- M + r^{2} + 16 \pi \int r[L+E^2 L_{,_{F}}]dr\right),
\label{FQ}
\end{eqnarray}
\begin{eqnarray}
N=\left(1-\frac{8 \pi
\omega^2(L+E^2L_{,_{F}})}{(1-\omega^2)}\right)^{-1}
\label{NthetaQ}
\end{eqnarray}
and
\begin{eqnarray}\label{Nthetaelec}
N^{\theta}=\frac{\omega \left(-M+16 \pi \int r[L+E^2 L_{,_{F}}]dr
\right)}{(1-\omega ^{2})R^{2}}.
\end{eqnarray}
The Maxwell tensor $F_{_{M}}=E \, dr \, \wedge \,d \tilde{t}$
transforms into
\begin{eqnarray}
F_{_{M}}=\frac{E}{\sqrt{1-\omega^2}} \, dr \, \wedge \,(dt-\omega
d \theta).
\end{eqnarray}
Due to this rotation we have now a magnetic component. Taking into
account the eq.~(\ref{N}) we have for the pseudoscalar magnetic
field, in the gauge~(\ref{metrica rotante}), the expression
\begin{eqnarray} B=\frac{E \, \omega \,
R}{\sqrt{1-\omega^2}\left(r-\frac{8 \pi
\omega^2r(L+E^2L_{,_{F}})}{(1-\omega^2)}\right)}.
\end{eqnarray}
The linear case is obtained when $L(F)=F=-E^2/2$ and all these
formulas became the equations (69)-(72) and (75) of the
Ref.\cite{Teitelboim1}.

Note that if we set $L(F)=0$ one gets
from~(\ref{felectrica})-(\ref{Nthetaelec}) the following metric
\begin{eqnarray}
ds^2=-\left(r^2-\frac{Mr^2}{r^2(1-\omega^2)+\omega^2M}
\right)dt^2+\frac{dr^2}{-M+r^2} \nonumber \\
+\left(r^2+\frac{\omega^2M}{1-\omega^2} \right)\left(d
\theta-\frac{\omega M}{r^2(1-\omega^2)+\omega^2M}dt \right)^2,
\end{eqnarray}
or equivalently
\begin{eqnarray}
ds^2=-\left(-\frac{M}{1-\omega^2}+r^2
\right)dt^2+\frac{dr^2}{-M+r^2} \nonumber \\ -\frac{2 \omega
M}{1-\omega^2}dt d\theta+\left(r^2+\frac{\omega^2 M}{1-\omega^2}
\right)d\theta^2.
\end{eqnarray}
This is the uncharged rotating BTZ black hole written in an
unusual gauge. To write this metric in the usual gauge we apply
the following radial coordinate transformation
\begin{eqnarray}
\rho^2=r^2+\frac{\omega^2 M}{1-\omega^2}
\end{eqnarray}
and we obtain
\begin{eqnarray}
ds^2=-(-\tilde{M}+\rho^2)dt^2+\frac{dr^2}{-\tilde{M}+\rho^2+J^2/(4\rho^2)}
\nonumber
\\ - J dt d\theta-\rho^2 d 
\theta^2=-\left(-\tilde{M}+\rho^2+\frac{J^2}{4\rho^2}
\right)dt^2 \nonumber \\
+\left(-\tilde{M}+\rho^2+\frac{J^2}{4\rho^2}
\right)^{-1}dr^2+\rho^2\left(\frac{J}{2 \rho^2}dt+d \theta
\right)^2,
\end{eqnarray}
where
\begin{eqnarray}\label{JyM}
J=\frac{2 \omega M}{1-\omega^2}, \,\,
\tilde{M}=\frac{M(1+\omega^2)}{1-\omega^2}.
\end{eqnarray}
So, the nonlinear electric rotating
solution~(\ref{felectrica})-(\ref{Nthetaelec}) reduces to the
uncharged rotating BTZ solution when the electric field vanishes.

An interesting example is obtained by application of this
procedure to the nonlinear regular black hole~(\ref{regular B-H}).
In this case the new rotating nonlinear electric spacetime is
given by the metric~(\ref{metrica rotante}), where $N^0= N G$ and
\begin{eqnarray}
R^2={\frac {{ r}^{2}-{\omega}^{2} \left( -M+{r}^{2}-q^2_e\ln
\left( {r}^{2}+{a}^{ 2} \right)  \right) }{1-{\omega}^{2}}}
\end{eqnarray}
\begin{eqnarray}
G^2= \frac{\left(r-{\omega}^{2} \left( r-{\frac {{q}^{2}_e
r}{{r}^{2}+{a}^{2}}}
  \right) \right)^2}{{r}^{2}-{\omega}^{2} \left( 
-M+{r}^{2}-{q}^{2}_e\ln  \left( {r}^{2}+{a}^
{2} \right)  \right)(1-\omega^2) } \times \nonumber \\ \nonumber
\\ (-M+{r}^{2}-{q}^{2}_e \ln  \left( {r}^{2}+{a}^{2} \right)),
\,\,\,\,\,\,\,\,\,\,\,\,\,\,\,\, {}^{}
\end{eqnarray}
\begin{eqnarray}
N={\frac { \left( {r}^{2}+{a}^{2} \right)  \left( -1+{\omega}^{2}
  \right) }{-{r}^{2}-{a}^{2}+{\omega}^{2}{r}^{2}+{\omega}^{2}{a}^{2}-{
\omega}^{2}{q}^{2}_e}},
\end{eqnarray}
and
\begin{eqnarray}
N^{\theta}={\frac {\omega\, \left( -M-{q}^{2}_e \ln  \left(
{r}^{2}+{a}^{2} \right) \right) }{{r}^{2}-{\omega}^{2} \left(
-M+{r}^{2}-{q}^{2}_e\ln  \left( { r}^{2}+{a}^{2} \right)
\right)}}.
\end{eqnarray}
This metric may be written also as
\begin{eqnarray}\label{metrica rotanttte}
ds^2= -{\frac {{q}^{2}_e \ln  \left( {r}^{2}+{a}^{2} \right)
+{\omega}^{2}{r}^{ 2}-{r}^{2}+M}{-1+{\omega}^{2}}} dt^2 \nonumber
\\ +{\frac {{{\it dr}}^{2}}{-M+{r}^{2}-{q}^{2}_e\ln  \left( {r}^{2}+{a}^{2}
\right) }} \nonumber \\ +{\frac {2\omega\, \left( -M-{q}^{2}_e \ln
\left( {r}^{2}+{a}^{2} \right)  \right) }{1-{\omega}^{2}}} \, dt
d\theta \nonumber \\ {\frac {{r}^{2}-{\omega}^{2} \left(
-M+{r}^{2}-{q}^{2}_e \ln  \left( {r}^ {2}+{a}^{2} \right)  \right)
}{1-{\omega}^{2}}} \, d \theta^2.
\end{eqnarray}

Other example is the rotating electric Coulomb-like spacetime,
which has the following form
\begin{eqnarray}
R^2=\frac{{r}^{2}-{\omega}^{2} \left( -M+{r}^{2}+{\frac
{{4q}^{2}_e}{3r }} \right)}{  \left( 1-{\omega}^{2} \right)},
\end{eqnarray}
\begin{eqnarray}
G^2= \nonumber \\ {\frac { \left(
-3\,{r}^{3}+3\,{\omega}^{2}{r}^{3}-2\,{\omega}^{2 }{q}^{2}_e
\right) ^{2} \left( -3\,Mr+3\,{r}^{3}+4\,{q}^{2}_e \right) }{{9r
}^{4} \left(
-3\,{r}^{3}-3\,{\omega}^{2}Mr+3\,{\omega}^{2}{r}^{3}+4\,{
\omega}^{2}{q}^{2}_e \right)  \left( -1+{\omega}^{2} \right) }},
\end{eqnarray}
\begin{eqnarray}
N={\frac {3{r}^{3} \left( -1+{\omega}^{2} \right)
}{-3\,{r}^{3}+3\,{ \omega}^{2}{r}^{3}-2\,{\omega}^{2}{q}^{2}_e}},
\end{eqnarray}
and
\begin{eqnarray}
N^{\theta}={\frac { \left( 3\,Mr-4\,{q}^{2}_e \right)
\omega}{-3\,{r}^{3}-3\,{
\omega}^{2}Mr+3\,{\omega}^{2}{r}^{3}+4\,{\omega}^{2}{q}^{2}_e}}.
\end{eqnarray}
This metric may be written also as
\begin{eqnarray} \label{Coulombiano magnetico}
ds^2=-\frac
{3\,Mr+3\,{\omega}^{2}{r}^{3}-3\,{r}^{3}-4\,{q}^{2}_e}{3r
  \left( -1+{\omega}^{2} \right)} \, dt^2 \nonumber \\
+ \frac{dr^2}{-M+r^2+4q^2_e/3r} + 2\,\omega\, \frac{-M+ {\frac
{{4q}^{2}_e}{3r}} }{  \left( 1-{ \omega}^{2} \right)} dt d\theta
\nonumber \\ +\frac{ {r}^{2}-{\omega}^{2} \left( -M+{r}^{2}+{\frac
{{4q}^{2}_e}{3r}} \right)} {\left( 1-{\omega}^{2} \right)}
d\theta^2.
\end{eqnarray}

\subsection{Stationary magnetic nonlinear spacetimes}
Now we shall consider the magnetostatic nonlinear spacetimes. In
order to write the magnetic rotating spacetime in linear
electrodynamics the authors of~\cite{Cataldo5} have applied this
kind of transformations. However, the properties of this spacetime
were not studied.  The extension to consider and study the
Einstein-Maxwell rotating magnetic solution has been derived in
Ref.~\cite{Lemos}. The angular momentum also is added
to~(\ref{HW}) through a rotational Lorentz boost, as we made
before for the electric extension. It is interesting to note that
the authors of Ref.~\cite{Lemos} interpret the static magnetic
field~(\ref{HW}) as being composed by a system of two symmetric
and superposed electric charges. In this case one of the electric
charges is at rest and the other is spinning. Thus there is no
electric field since the total electric charge is zero and the
magnetic field is generated by the angular electric current.

For obtaining the Einstein-nonlinear Maxwell rotating magnetic
solution we shall apply the transformation~(\ref{boost}) to the
magnetostatic solution~(\ref{solucionprefinal}). Then we obtain
the metric
\begin{eqnarray} \label{metr mag rot}
ds^2&=& \frac{r^2-\omega^2(-M+r^2+16 \pi \int r(L-B^2
L_{_{F}})dr)}{1-\omega^2} \, dt^2 \nonumber \\
&-&\frac{dr^2}{-M+r^2+16 \pi \int r(L-B^2 L_{_{F}})dr)} \nonumber
\\ &+&\frac{2 \omega (-M+16 \pi \int r(L-B^2
L_{_{F}})dr))}{1-\omega^2} \, dt\, d\theta \nonumber \\
&-&\frac{-M+r^2+16 \pi \int r(L-B^2
L_{_{F}})dr)-r^2\omega^2}{1-\omega^2} d \theta^2. \nonumber \\
\end{eqnarray}
The Maxwell tensor $F_{_{M}}=B \, dr \wedge d \tilde{\theta}$
transforms into
\begin{eqnarray} \label{maxwell rotacional magnetico}
F_{_{M}}=\frac{B}{\sqrt{1-\omega^2}} \, dr \wedge (d \theta-\omega
dt).
\end{eqnarray}
In this case due to the rotation we have now an electric
component.

The metric~(\ref{metr mag rot}) can be rewritten as
\begin{eqnarray} \label{metrica rotante magnetica}
ds^2=-R^2(r)(dt+N^{\theta}(r)\, d\theta)^2+\frac{dR^2}{G^2(r)}
+(N^0(r))^2 \, d\theta^2, \nonumber \\
\end{eqnarray}
where $N^0(r)=N(r)G(r)$ and
\begin{eqnarray}
R^2=r^2+\frac{\omega^2}{1-\omega^2} \left(M - 16 \pi \int r[L-B^2
L_{,_{F}}]dr \right), \label{MR2Q}
\end{eqnarray}
\begin{eqnarray}
G^2=\frac{\left(r-\frac{8 \omega^2\pi r (L-B^2
L_{,_{F}})}{1-\omega^2} \right)^2}{R^2} \nonumber \\ \times
\left(- M + r^{2} + 16 \pi \int r[L-B^2 L_{,_{F}}]dr\right),
\label{MFQ}
\end{eqnarray}
\begin{eqnarray}
N=\left(1-\frac{8 \pi
\omega^2(L-B^2L_{,_{F}})}{(1-\omega^2)}\right)^{-1},
\label{MNthetaQ}
\end{eqnarray}
\begin{eqnarray}\label{Nthetamag}
N^{\theta}=\frac{\omega \left(-M+16 \pi \int r[L-B^2 L_{,_{F}}]dr
\right)}{(1-\omega ^{2})R^{2}},
\end{eqnarray}
and $dr/dR=NR/r$. Considering the eq.~(\ref{maxwell rotacional
magnetico}) we obtain for the electric component
\begin{eqnarray}
E=\frac{B \, \omega \, R}{\sqrt{1-\omega^2}\left(r-\frac{8 \pi
\omega^2r(L-B^2L_{,_{F}})}{(1-\omega^2)}\right)}.
\end{eqnarray}
Comparing the form of the metric~(\ref{metrica rotante}) with the
line element~(\ref{metrica rotante magnetica}) and the
expresions~(\ref{R2Q})-(\ref{Nthetaelec}) with the
eqs.~(\ref{MR2Q})-(\ref{Nthetamag}) we find that both solutions
are connected by the same duality mapping, which in this case may
be written as
\begin{eqnarray}\label{dual map}
dt \longrightarrow i d \theta, d \theta \longrightarrow i dt, E
\longrightarrow iB.
\end{eqnarray}
Thus, for example, the magnetic rotating counterparts of the
solutions~(\ref{regular magnetic}) and~(\ref{magnetic coulomb})
may be obtained applying the duality mapping~(\ref{dual map}) to
the eqs.~(\ref{metrica rotanttte}) and~(\ref{Coulombiano
magnetico}) respectively.

In conclusion these electric and magnetic rotating metrics have
both locally static spacetimes which are connected by a duality
mapping. Then the rotating spacetimes are also connected by the
same duality mapping.

\section{Some properties of spinning nonlinear three-dimensional
spacetimes} Both the static and rotating solutions are
asymptotically anti-de Sitter. Then it is interesting, for
example, to calculate the angular momentum and mass of the
rotating solutions.

\subsection{Spinning electric solutions}
To identify the angular momentum in the metric of the
form~(\ref{metrica rotante}) we can use the quasilocal
formalism~\cite{Brown1,Brown2}. In this case the angular momentum
$j(r)$ at a radial boundary $r$ is given by~\cite{Chan}
\begin{eqnarray*}
j(r)=\frac{G N^{\theta^{\,\prime}}R^3}{N^0 \, R^{\,\prime}},
\end{eqnarray*}
where the prime denotes an ordinary derivative with respect to
$r$. Then $J$ is defined as $J=j(\infty)$. In view that in our
case $N^0(r)=N(r) G(r)$ we have that
\begin{eqnarray}
j(r)=\frac{N^{\theta^{\,\prime}}R^3}{N \, R^{\, \prime}}.
\end{eqnarray}
For the rotating nonlinear electric
spacetime~(\ref{felectrica})-(\ref{Nthetaelec}) we have that the
angular momentum at a radial boundary $r$ is given by
\begin{eqnarray}
j(r)=\frac{2 \omega}{1-\omega^2} \times
\,\,\,\,\,\,\,\,\,\,\,\,\,\,\,\,\,\,\,\,\,\,\,\,\,\,\,\,\,\, {}^{}
\nonumber
\\ \left[ M+8\pi r^2(L+E^2L_{_{F}})-16\pi\int r(L+E^2L_{_{F}})dr
\right].
\end{eqnarray}
 From the last equation we obtain for the uncharged rotating BTZ
black hole ($L(F)=0$) that $j(r)=2 \omega M/(1-\omega^2)$, i.e.
the first relation of eq.~(\ref{JyM}).

In general the behavior of $j(r)$ at infinity depends on the
considered electrodynamics. For the linear case
$L=-F/(4\pi)=E^2/(8\pi)$. Then we can write
$L+E^2L_{_{F}}=-q^2/(8\pi r^2)$ and then
\begin{eqnarray}
j(r)=\frac{2\omega}{1-\omega^2} \left(M-q^2_e+2q^2_e \ln r
\right).
\end{eqnarray}
At infinity this expression diverges, as it is well known. This
intrinsic logarithmic divergence is due to the fact that the
electrostatic potential is logarithmic in the linear case.

\begin{table}
\caption{Expressions for the angular momentum at a radial boundary
$r$, $j(r)$, and angular momentum, $J$, for the rotating nonlinear
electric Born-Infeld, regular and Coulomb-like black holes.}
\begin{tabular}{|c|c|c|}
\hline  Rotating        &   &
\\ Spacetimes & j(r) & J
\\ &    &
\\ \hline &&
\\
   from (\ref{Born-Infeld B-H}) & $\frac{2\omega}{1-\omega^2} 
\left(M+2q^2_e \ln \left (
r+\sqrt{r^2+\frac{q^2_e}{b^2}} \right)\right)$ & $\infty$
\\ && \\ \hline
&&\\ (\ref{metrica rotanttte}) & $\frac{2\omega}{1-\omega^2}
\left(M+q^2_e \ln \, (r^2+a^2)-\frac{q^2_e r^2}{r^2+a^2} \right) $
& $\infty$
\\ && \\ \hline
&& \\ (\ref{Coulombiano magnetico}) & $\frac{2\omega}{1-\omega^2}
\left(M-\frac{2q^2_e}{r} \right)$ & $\frac{2\omega}{1-\omega^2}M$
\\ && \\ \hline
\end{tabular}
\end{table}

The behavior of the angular momentum $j(r)$ at a radial boundary
$r$ and its behavior at infinity ($J$) for nonlinear electric
rotating black holes generated by the boost~(\ref{boostt}) from
the nonlinear electrostatic spacetimes~(\ref{Born-Infeld B-H}),
(\ref{regular B-H}) and~(\ref{Coulomb-like elec}) are summarized
on Table I. The nonlinear Born-Infeld electric black
hole~(\ref{Born-Infeld B-H}) and  the regular black
hole~(\ref{regular B-H}) have an angular momentum, which diverges
at infinity. More interesting is the black hole with Coulomb-like
field~(\ref{Coulomb-like elec}). At infinity this rotating
spacetime has the same angular momentum $J$ of the uncharged
rotating BTZ solution.

In addition, we also shall calculate the mass of solutions. One
can use the quasilocal mass formula developed
in~\cite{Brown1,Brown2} (see also~\cite{Chan}). In this case, from
eq.~(\ref{metrica rotante}), we have that the quasilocal energy
$E(r)$ and the quasilocal mass $M(r)$ at a radial boundary $r$ can
be shown to be respectively
\begin{eqnarray}\label{quasilocal energy}
E(r)=2 \left(f_{ref} \frac{dR_{ref}}{dr}-f(r) \frac{dR}{dr}
\right),
\end{eqnarray}
\begin{eqnarray}\label{quasilocal mass}
M(r)=2NGE(r)-j(r) N^{\theta},
\end{eqnarray}
where $f^2(r)=(dR/dr)^{-2} \, G^2$ and $f_{ref}$ and $R_{ref}$ are
the background structural functions which determine the zero of
energy. The background metric can be obtained simply by setting
constant of integration of a particular solution to some special
value, determining this way the reference spacetime. We set $q=0$
and $M=0$ in our rotating solutions, arriving at the background
spacetime, which corresponds to an asymptotic vacuum anti-de
Sitter metric. Then $L_0 dR_0/dr=r/l$ ( with $l=1$). The same
background was used in other works for the calculation of the
quasilocal mass of some spinning solutions (see for
example~\cite{Brown2,Chan}). The quasilocal mass at spatial
infinity is defined as
\begin{eqnarray*}
\tilde{M}=m(\infty)
\end{eqnarray*}

The mass and angular momentum may be also obtained applying the
formalism of Regge and Teitelboim~\cite{Regge}. For example, for
the canonical form of the metric~(\ref{metrica rotante}) the mass
is given by (the details can be found in~\cite{Lemos})
\begin{eqnarray}
M(r)=\nu [-R_{,r}(f^2-f_{ref}^2)+(f^2)_{,r}(R-R_{ref}) \nonumber
\\ -2f^2(R_{,r}-R_{ref_,r})],
\end{eqnarray}
where $f^2=(dR/dr)^{-2} \, G^2$, $f_{ref}^2$ and $R_{ref}$ are
structural functions of a background reference spacetime and $\nu$
is a constant factor, which come from the fact that, due the angle
deficit, the integration over azimuthal angle $\theta$ is between
$0$ and $2 \pi \nu$ (see~\cite{Lemos}) .

It can be shown that the quasilocal mass at spatial infinity for
the charged spinning BTZ solution and the rotating regular black
hole is given by
\begin{eqnarray}\label{la masa}
\tilde{M}=\frac{1+\omega^2}{1-\omega^2} \, M- \frac{2 \omega^2
q^2_e}{1-\omega^2}+\frac{2q^2_e}{1-\omega^2} \ln r.
\end{eqnarray}
This coincidence takes place because the electric
solutions~(\ref{BTZ}) and~(\ref{regular B-H}) have the same
behavior at spatial infinity in static and stationary cases.

For the rotating electric Coulomb-like black hole the quasilocal
mass at spatial infinity does not diverges. In this case we have
$\displaystyle \tilde{M}=\frac{1+\omega^2}{1-\omega^2} \, M$, i.e.
the same value of the uncharged rotating BTZ black hole.

The behavior of the quasilocal mass at infinity, $\tilde{M}$, for
nonlinear electric rotating black holes generated
from~(\ref{Born-Infeld B-H}), (\ref{regular B-H})
and~(\ref{Coulomb-like elec}) are summarized on Table II.

\begin{table}
\caption{Expressions of the quasilocal mass at spatial infinity
for the rotating nonlinear electric Born-Infeld, regular and
Coulomb-like black holes.}
\begin{tabular}{|c|c|c|}
\hline 
Rotating&& \\  Spacetimes & $\tilde{M}$ &   Behavior at $\infty$
\\ & &
\\ \hline &&
\\
from (\ref{Born-Infeld B-H}) & $\frac{1+\omega^2}{1-\omega^2}
\left(M + q^2_e \ln r^2+q^2_e  \ln 4 \right)+q^2_e$ & $\infty$
\\ && \\ \hline
&&\\ (\ref{metrica rotanttte}) & $\frac{1+\omega^2}{1-\omega^2} \,
M- \frac{2 \omega^2 q^2_e}{1-\omega^2}+\frac{2q^2_e}{1-\omega^2}
\ln
  r$ & $\infty$
\\ && \\ \hline
&& \\ (\ref{Coulombiano magnetico}) &
$\frac{1+\omega^2}{1-\omega^2} \, M$ &
$\frac{1+\omega^2}{1-\omega^2} \, M$
\\ && \\ \hline
\end{tabular}
\end{table}
Thus $\tilde{M}$ diverges logarithmically for the Born-Infeld and
regular black holes, similar to the situation in the angular
momentum.

\subsection{Spinning magnetic solutions}
For the magnetic solution~(\ref{metr mag rot}) we can write the
angular momentum and mass in the form
\begin{eqnarray}
j(r)=\frac{f(N^0-R^2N^{\theta^{^2}})^2}{R
N^0}\left(\frac{N^{\theta}R^2}{R^2N^{\theta^{^2}}-N^{0^{2}}}\right)^{\prime}
\end{eqnarray}
and
\begin{eqnarray}
M(r)=\nu \left[-\sqrt{N^{0^{^2}}-R^2N^{\theta^{^2}}}
\left(f^2-f^2_{ref} \right) \right.\nonumber \\ \left.
(f^2)_{,r}\left(
\sqrt{N^{0^{^2}}-R^2N^{\theta^{^2}}}-K_{ref}\right) \right.
\nonumber \\ \left. -2 f^2 \left(\left
(\sqrt{N^{0^{^2}}-R^2N^{\theta^{^2}}} \right)_{,r} -K_{ref
\,,r}\right) \right],
\end{eqnarray}
respectively, where the structural functions $f$, $R$, $N^0$ and
$N^{\theta}$ are given by eqs~(\ref{metrica rotante
magnetica})-(\ref{Nthetamag}).

The behavior of the angular momentum and mass at a radial boundary
$r$ and its behavior at infinity for rotating nonlinear magnetic
spacetimes generated from the static solutions~(\ref{magnetic
B-I}), (\ref{regular magnetic}) and~(\ref{magnetic coulomb}) are
summarized on the Tables III and IV respectively.
\begin{table}
\caption{Expressions for the angular momentum at a radial boundary
$r$, $j(r)$, and angular momentum, $J$, for the rotating nonlinear
magnetic spacetimes generated from the static
solutions~(\ref{magnetic B-I}), (\ref{regular magnetic})
and~(\ref{magnetic coulomb}).}
\begin{tabular}{|c|c|c|}
\hline   Rotating                    &          &        \\
          Spacetimes                  & j(r)     & J       \\
generated                        &          &         \\ from & &
\\

\hline & &
\\ (\ref{magnetic B-I}) & $\frac{2\omega}{\omega^2-1}
\left(M-2q^2_m \ln \left ( r+\sqrt{r^2-\frac{q^2_m}{b^2}}
\right)\right)$ & $\infty$
\\ && \\
\hline
&&\\ (\ref{regular magnetic}) & $\frac{2\omega}{\omega^2-1}
\left(M-q^2_m \ln \, (r^2+a^2)+\frac{q^2_m r^2}{r^2+a^2} \right) $
& $\infty$
\\ && \\ \hline
&& \\ (\ref{magnetic coulomb}) & $\frac{2\omega}{\omega^2-1}
\left(M+\frac{2q^2_m}{r} \right)$ & $\frac{2\omega}{\omega^2-1} \,
M$
\\ && \\ \hline
\end{tabular}
\end{table}
\begin{table}
\caption{Expressions of the mass at spatial infinity for the
rotating nonlinear magnetic spacetimes generated from the static
solutions~(\ref{magnetic B-I}), (\ref{regular magnetic})
and~(\ref{magnetic coulomb}).}
\begin{tabular}{|c|c|c|}
\hline Rotating              &                      & \\
Spacetimes   & $\tilde{M}$ & Behavior at $\infty$      \\
generated                        &          &         \\ from & &
\\
\hline && \\
   (\ref{magnetic B-I}) & $\frac{\omega^2+1}{\omega^2-1} \left( M - q^2_m 
\ln r^2-q^2_m
\ln 4 \right)-q^2$ & $\infty$
\\ && \\ \hline
&&\\ (\ref{regular magnetic}) & $\frac{\omega^2+1}{\omega^2-1} \,
M+ \frac{2 q^2_m}{\omega^2-1}-\frac{\omega^2+1}{\omega^2-1}q^2_m
\ln r^2$ & $\infty$
\\ && \\ \hline
&& \\ (\ref{magnetic coulomb}) & $\frac{\omega^2+1}{\omega^2-1} \,
M$ & $\frac{\omega^2+1}{\omega^2-1} \, M$
\\ && \\ \hline
\end{tabular}
\end{table}

Comparing the expressions for the angular momentum of rotating
magnetic solutions (given in table III) with the electric one
(given in the table I), we conclude that the angular momentum of
the electric rotating spacetimes and the magnetic rotating one are
connected by the duality mapping~(\ref{dual map}).

The divergence of the mass may be handled as
follows~\cite{Teitelboim1,Lemos}.
We can rewrite, for example, the eq.~(\ref{la masa}) in the
following form
\begin{eqnarray}\label{divergence Mass}
\tilde{M}=\frac{1+\omega^2}{1-\omega^2} \, M - \frac{2 \omega^2
q^2_e}{1-\omega^2} + Div_{_{M}}(r),
\end{eqnarray}
where $Div_{_{M}}(r)$ is the logarithmic term, which diverges if
$r \longrightarrow \infty$. Now, one encloses the system in a
boundary of large radius $r_0$. Then, one sums and subtracts
$Div_{_{M}}(r_0)$ to~(\ref{divergence Mass}) so that this mass is
now written as
\begin{eqnarray} \label{divergence Mass1}
\tilde{M}=M(r_0)+ \left( Div_{_{M}}(r)-Div_{_{M}}(r_0) \right),
\end{eqnarray}
where
\begin{eqnarray} \label{divergence Mass11}
M(r_0)=M_0+Div_{_{M}}(r_0).
\end{eqnarray}
The second and third terms of~(\ref{divergence Mass1}) vanish when
$r \longrightarrow r_0$. One might call $M(r_0)$ the energy within
the radius $r_0$. This energy differs from $M_0$ by
$-Div_{_{M}}(r_0)$ which may be interpreted as the electromagnetic
energy outside $r_0$ up to an infinite constant which is absorbed
in $M(r_0)$. Then from~(\ref{divergence Mass11}) we have that
\begin{eqnarray}\label{divergence Mass115}
M_0=M(r_0)-Div_{_{M}}(r_0).
\end{eqnarray}
and this sum is independent of $r_0$, finite and equal to the
total mass.

Thus in practice the treatment of the mass divergence amounts to
forgetting about $r_0$ and takes as zero the asymptotic limit
\begin{eqnarray*}
\lim_{r \rightarrow \infty} \, Div_{_{M}}(r)=0
\end{eqnarray*}
in eq.~(\ref{divergence Mass}).

The divergence on the angular momentum can be treated in a similar
way as the mass divergence.


\section{Acknowledgements}
I thank Carol Mu\~noz for typing this manuscript. This work was
supported by CONICYT through Grant FONDECYT N$^0$ 1010485 and by
Direcci\'on de Promoci\'on y Desarrollo de la Universidad del
B\'\i o-B\'\i o.

\end{document}